\documentclass[10pt,journal,letter,twoside,twocolumn]{IEEEtran}
\usepackage{subfigure}
\usepackage{longtable}
\usepackage{stfloats, balance, epsfig}
\usepackage{subfigure}
\usepackage{colortbl}
\usepackage{graphicx}
\usepackage{subfigure}
\usepackage{multirow}
\usepackage{lscape}

\usepackage{multirow}
\usepackage{setspace}

\usepackage{graphicx}  
\usepackage{url}       
\usepackage{graphics}

\usepackage{amsmath}   
\usepackage{cite}
\DeclareMathOperator*{\argmax}{argmax}
\DeclareMathOperator*{\argmin}{argmin}
\usepackage{amsfonts,amssymb}
\usepackage{booktabs}
\usepackage{threeparttable}
\usepackage{stfloats}
\usepackage{xcolor}
\usepackage{cases}
\usepackage{multirow}
\usepackage{multicol}
\usepackage{subeqnarray}
\usepackage[ruled,vlined]{algorithm2e}
\usepackage{algpseudocode}

\usepackage{xcolor}
\usepackage{cite}

\usepackage{geometry}
\usepackage{enumerate}
\makeatletter
\newenvironment{tablehere}
{\def\@captype{table}}
{}

\makeatother

\newcommand{\ls}[1]
    {\dimen0=\fontdimen6\the\font
     \lineskip=#1\dimen0
     \advance\lineskip.5\fontdimen5\the\font
     \advance\lineskip-\dimen0
     \lineskiplimit=.9\lineskip
     \baselineskip=\lineskip
     \advance\baselineskip\dimen0
     \normallineskip\lineskip
     \normallineskiplimit\lineskiplimit
     \normalbaselineskip\baselineskip
     \ignorespaces
    }
\newcommand\figurecaption{\def\@captype{figure}\caption}
\newcommand\tablecaption{\def\@captype{table}\caption}

\begin{document}

\makeatletter
\newcommand{\rmnum}[1]{\romannumeral #1}
\newcommand{\Rmnum}[1]{\expandafter\@slowromancap\romannumeral #1@}
\makeatother

\title{Mixed-Timescale Per-Group Hybrid Precoding for Multiuser Massive MIMO Systems}
\author{\small\IEEEauthorblockN{Yinglei Teng, Min Wei, An Liu, Vincent Lau, Yong Zhang}
}
\author{
Yinglei Teng, Min Wei, An Liu,  \textit{Senior Member, IEEE}, Vincent Lau, \textit{Fellow, IEEE}\thanks{This work was supported in part by the NSFC under Grant NO. 61427801 and No. 61771072, and BNSF under Grant No. L171011. Yinglei Teng, Min Wei and Yong Zhang are with the Beijing Key Laboratory of Work Safety Intelligent Monitoring, BUPT, China (e-mail: lilytengtt@gmail.com; {weimin2016140309, yongzhang}@bupt.edu.cn). An Liu is with Zhejiang University (e-mail: wendaolstr@gmail.com). Vincent Lau is with the department of ECE, HKUST, HK (e-mail: eeknlau@ece.ust.hk).}, and Yong Zhang
}

\maketitle
\vspace{-2ex}
\begin{abstract}
Considering the expensive radio frequency (RF) chain, huge training overhead and feedback burden issues in massive MIMO, in this letter, we propose a mixed-timescale per-group hybrid precoding (MPHP) scheme under an adaptive partially-connected RF precoding structure (PRPS), where the RF precoder is implemented using an adaptive connection network (ACN) and $M$ analog phase shifters (APSs), where $M$ is the number of antennas at the base station (BS). Exploiting the mixed time stage channel state information (CSI) structure, the joint-design of ACN and APSs is formulated as a statistical signal-to-leakage-and-noise ratio (SSLNR) maximization problem, and a heuristic group RF precoding (GRFP) algorithm is proposed to provide a near-optimal solution. Simulation results show that the proposed design advances at better energy efficiency (EE) and lower hardware cost, CSI signaling overhead and computational complexity than the conventional hybrid precoding (HP) schemes.
\end{abstract}



\vspace{-1mm}
\section{Introduction}\label{sec:intro}
\IEEEPARstart{M}{assive} multiple-input multiple-output (MIMO), in which the base station (BS) is equipped with $M\gg 1$ antenns, has been a pivotal technology for the forthcoming fifth-generation (5G) cellular system \cite{7414384}. However, employing the conventional digital precoding in massive MIMO requires a large number of radio frequency (RF) chains, which induces huge hardware cost and power consumption. To overcome this problem, the concept of hybrid precoding (HP), consisting of both digital and analog precoding, is proposed in\cite{2004:Zhang-phase}.\\
\indent
The architecture of HP depends heavily on the antenna structure. Existing HP schemes are primarily based on the \textit{fully-connected RF precoding  structure} (FRPS)\cite{Liu2014Phase}, \cite{Ayach2014Spatially}, where each RF chain is connected to every antenna through analog phase shifters (APSs) and RF adders. The requirement for a large number of APSs and RF adders increases the hardware cost and power consumption. To address this problem, a \textit{partially-connected RF precoding structure} (PRPS) is proposed in \cite{Molisch2016Hybrid}, \cite{6824962} such that each RF chain is only connected to a subset of antennas. Under such a PRPS, \cite{Zhu2016Adaptive} proposes an adaptive hybrid precoding (AHP) scheme implemented by an adaptive connection network (ACN) and $M$ APSs, where both the RF precoder and baseband precoder are adaptive to the real-time full channel state information (CSI). However, it is very difficult to obtain such full CSI in frequency division duplexing (FDD) massive MIMO systems due to the limited number of pilots. In \cite{Liu2014Phase} and \cite{7572969}, a mixed-timescale HP scheme is proposed, where the RF precoder is adaptive to the second order channel statistics, and the baseband precoder is adaptive to the reduced-dimension effective CSI. Such a mixed-timescale design addresses the issue of insufficient pilot symbols in the training stage. However, it is studied only under the FRPS, which has high hardware complexity and cost.\\
\indent
In this letter, we propose a mixed-timescale per-group hybrid precoding (MPHP) scheme under an adaptive PRPS. In the long term, the RF precoder is optimized through the maximization of a statistical signal-to-leakage-and-noise ratio (SSLNR) and a low complexity group RF precoding (GRFP) algorithm is proposed to provide a near-optimal solution. In the short term, the zero-forcing (ZF) precoder is employed adaptive to the low-dimensional real-time effective CSI. Simulations are presented from the aspects of average rate, energy efficiency (EE) and fairness. Analyzed the feedback burden and complexity, the proposed design can achieve better EE but with significant reduction of the hardware cost, CSI signaling overhead and computational complexity.
\vspace{-3mm}
\section{System Model}\label{SysMod}
\subsection{Multi-user Massive MIMO System with an adaptive PRPS}
Consider a downlink (DL) multi-user massive MIMO system with limited RF chains. The system consists of a BS with $M\gg 1$ antennas, $L$ transmit RF chains and $K$ single-antenna users, where $K\leq L<M$. Let $\mathcal{M}$ and $\mathcal{L}$ denote the antenna set and RF chain set, respectively, where $\vert\mathcal{M}\vert=M$ and $\vert\mathcal{L}\vert=L$. \\
\indent
For clarity, we focus on the case when $L=K$. To support simultaneous downlink transmissions to the $K$ users with $L=K$ RF chains, we employ a two-stage precoding structure $\mathbf{Q}=\mathbf{FW}$ as illustrated in \textmd{Fig.}~\ref{fig:firure_1}, where $\mathbf{F}\in\mathbb{C}^{M\times L} $  is the RF precoder implemented by an ACN and $M$ APSs. Specifically, each antenna is only connected to one RF chain through one APS, and the $l$-th RF chain is connected to $N_l$ antennas through $N_l$ APSs, where $\sum_{l=1}^L N_l =M$. The connection between the RF chains and antennas can be dynamically adjusted using an ACN, which is essentially a programmable RF switch. The joint-design of the ACN and APSs is achieved by optimizing the RF precoder $\mathbf{F}$ based on the channel statistics, where the locations of non-zero elements in $\mathbf{F}$ determine the ACN (the connections between the antennas and the RF chains in the ACN) and the phases of the non-zero elements decide the phases of APSs. The baseband precoder $\mathbf{W}\in\mathbb{C}^{L\times K} $ is adaptive to the real-time effective CSI with respect to the $L$ RF chains. With MPHP, the received DL signal for all $K$ users can be represented as
\vspace{-2mm}
\begin{equation}\label{Eq:1}
\mathbf{y}=\mathbf{H}^H\mathbf{Q}\mathbf{P}^{\frac{1}{2}}\mathbf{s}
	+\mathbf{n}
	={\overline{\mathbf{H}}}\mathbf{W}\mathbf{P}^{\frac{1}{2}}\mathbf{s}
	+\mathbf{n},
\vspace{-1mm}
\end{equation}
\noindent
where $\mathbf{H}^H\in\mathbb{C}^{K\times M} $ denotes the DL channel matrix between the BS and all $K$ users, $\mathbf{H}=[\mathbf{h}_1,\dots,\mathbf{h}_k,\dots,\mathbf{h}_K]$, $\mathbf{h}_k\in \mathbb{C}^{M\times 1}$ is the channel vector of user $k$, $\mathbf{P}\in\mathbb{C}^{K\times K}$ is a diagonal matrix to maintain the total transmit power $P$, $\mathbf{s}=[s_1,\dots,s_K]^T$ is the data vector, $s_k\sim \mathcal{CN}(0,1)$ denotes the desired date symbol of user $k$, and $\mathbf{n}\sim \mathcal{CN}(0,\mathbf{I}_K)$ is the corresponding additive white Gaussian noise vector. Moreover, $\overline{\mathbf{H}}=\mathbf{H}^H\mathbf{F}\in\mathbb{C}^{K\times K}$ represents the effective downlink channel after RF precoding.
 \vspace{-1mm}
	\begin{center}
		\includegraphics[width=7.5cm,height=4.6cm]{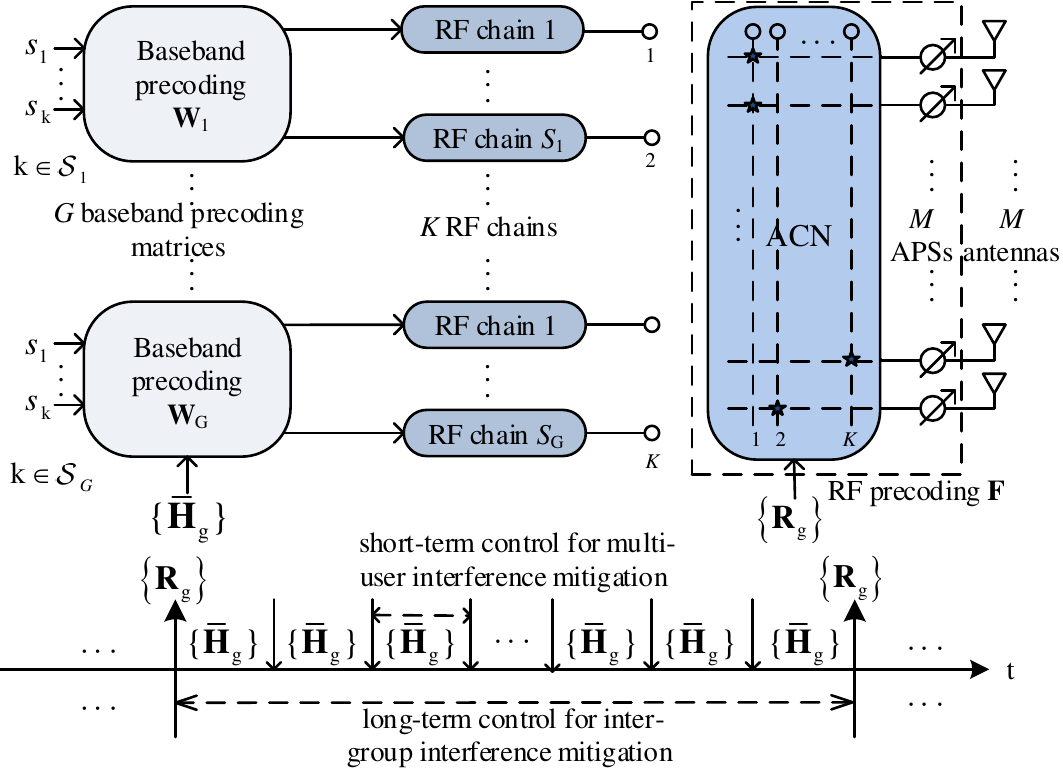}
		\makeatletter\def\@captype{figure}\makeatother
		\vspace{-2mm}
		\caption{The architecture of the proposed MPHP with the ACN and $M$ APSs.}\label{fig:firure_1}
	\end{center}
\vspace{-5mm}
\subsection{Spatial Channel Model and User Grouping}
We consider a block fading channel where the channel matrix remains constant within each block and changes at the boundary of the block according to a general distribution. For each user $k$, the spatial correlation matrix $\mathbf{R}_{k}\in \mathbb{C}^{M\times M}$ is defined as 
\vspace{-1mm}
\begin{equation}\label{Eq:6}
\mathbf{R}_{k}=\mathrm{E}[\mathbf{h}_k\mathbf{h}_k^H].
\vspace{-1mm}
\end{equation}
\indent
In practice, users with similar angle-of-departure (AoD) distributions can exhibit similar spatial correlation matrices. Users are partitioned into different groups by the similarity of $\mathbf{R}_{k}$ by a user grouping algorithm. Let $\mathcal{G}$ denote the group set and $\mathcal{S}_g$ be the user set of group $g$ with $\vert\mathcal{G}\vert=G$ and $\vert \mathcal{S}_g \vert=S_g$, where $S_g$ is the number of users in group $g$. After user grouping, the average spatial correlation matrix of group $g$ is $\mathbf{R}_g=\frac{1}{S_g}\sum_{k\in \mathcal{S}_g}\mathbf{R}_k, \forall g\in \mathcal{G}$.\\
\indent
Assume users are classified into $G$ groups by an appropriate grouping method\footnote{The proposed MPHP utilizes the user grouping algorithm in \cite{Xu2014User}.}, let $\mathcal{L}_g$ be the RF chain set of group $g$ with $\vert \mathcal{L}_g \vert=S_g$. Without loss of generality, we assume that the index sets of users and RF chains of the $g$-th group are given by $\mathcal{S}_g=\{\sum_{j=1}^{g-1}S_j+1,\ldots, \sum_{j=1}^{g}S_j\}$ and $\mathcal{L}_g=\{\sum_{j=1}^{g-1}S_j+1,\ldots, \sum_{j=1}^{g}S_j\}$, respectively. Then, the channel matrix can be written as $\mathbf{H}=[\mathbf{H}_1,\dots,\mathbf{H}_g,\dots,\mathbf{H}_G]$, where $\mathbf{H}_g\in\mathbb{C}^{M \times S_g}$ is the channel matrix of group $g$. Similarly, the RF and baseband precoding matrix can be denoted as $\mathbf{F}=[\mathbf{F}_1,\dots,\mathbf{F}_g,\dots,\mathbf{F}_G]$ and $\mathbf{W}=\mathrm{diag}(\mathbf{W}_1,\dots,\mathbf{W}_g,\dots,\mathbf{W}_G)$, respectively, where $\mathbf{F}_g=[\mathbf{f}_l]_{l\in \mathcal{L}_g}\in \mathbb{C}^{M\times S_g}$ is the RF precoding matrix for group $g$, and $\mathbf{W}_g=[\mathbf{w}_k]_{k\in \mathcal{S}_g}\in \mathbb{C}^{S_g\times S_g}$ with $\Vert \mathbf{w}_k\Vert=1$ is the baseband precoder for group $g$. Then, the signal received by the users in group $g$ can be written as
\vspace{-1.5mm}
\begin{equation}\label{Eq:2}
\mathbf{y}_{g}=\mathbf{H}_{g}^H \mathbf{F}_g\mathbf{W}_{g}\mathbf{P}_g^{\frac{1}{2}}\mathbf{s}_{g}+{\sum\nolimits_{g'\neq g}\mathbf{H}_{g}^H \mathbf{F}_{g'}\mathbf{W}_{g'}\mathbf{P}_{g'}^{\frac{1}{2}}\mathbf{s}_{g'}}+\mathbf{n}_{g},
\vspace{-2mm}
\end{equation}
where $\mathbf{s}_{g}=[s_k]_{k\in \mathcal{S}_g} \in \mathbb{C}^{S_g\times 1}$ is the data vector of the $g$-th group, $\mathbf{P}_g=\mathrm{diag}({p_{\sum_{j=1}^{g-1}S_j+1}},\dots,{p_k},\dots,{p_{\sum_{j=1}^{g}S_j}})$ with $p_k=\frac{{P}}{{K}{\Vert\mathbf{F}_g\mathbf{w}_k\Vert}^2},\forall k\in\mathcal{S}_g$ is a power normalization matrix of group $g$ to ensure that the total transmit power is $P$, and $\mathbf{n}_{g}\sim \mathcal{CN}(0,\mathbf{I}_{S_g})$ is the additive white Gaussian noise vector. The downlink signal-to-interference-plus-noise ratio (SINR) of user $k$ in group $g$ is given by
\vspace{-2mm}
\begin{equation}\label{Eq:3}
\mathrm{SINR}_{k}=\dfrac{p_k{\lvert \mathbf{h}_{k}^H \mathbf{F}_g\mathbf{w}_{k} \rvert}^2}{\sum\nolimits_{i\in \mathcal{S}_{g'}, g'\neq g}p_i {\lvert \mathbf{h}_{k}^H \mathbf{F}_{g'}\mathbf{w}_{i}\rvert}^2 +1}.
\vspace{-1mm}
\end{equation}
Then, the corresponding achievable rate of user $k\in \mathcal{S}_g,\forall g \in \mathcal{G}$ can be represented as $R_{k}=\mathbb{E}_{\mathbf{H}}\{\log_2(1+\mathrm{SINR_{k}})\}$.
\vspace{-1mm}
\section{Proposed Hybrid Precoding Scheme}\label{ProHP}
\vspace{-1mm}
\subsection{Short-Term Baseband Precoding}
For a given RF precoding matrix $\mathbf{F}_g$, a simple ZF precoder \cite{Liang2014Low} is employed at the baseband to eliminate the intra-group interference based on the knowledge of the low-dimensional effective CSI $\overline{\mathbf{H}}_{g}={\mathbf{H}}_{g}^H {\mathbf{F}}_{g}\in \mathbb{C}^{S_g\times S_g}$. Specifically, the baseband precoder $\mathbf{W}_g$ is obtained by normalizing each column of $\overline{\mathbf{H}}^{H}_{g}(\overline{\mathbf{H}}_{g} \overline{\mathbf{H}}^{H}_{g})^{-1}$. 
\vspace{-6mm}
\subsection{Optimization Formulation for Long-Term RF Precoding}
Instead of optimizing the SINR-related performance, which would lead to a difficult non-convex optimization problem, we focus on the signal-to-leakage-and-noise ratio (SLNR) related performance metric. SLNR has been widely used as a performance metric for low-complexity and near-optimal precoder design in conventional MIMO downlink systems. The SLNR for user $k$ in group $g$ is defined as\\
\vspace{-3mm}
\begin{equation}\label{66}
{\rm{SLNR}}_{k}^{} = \frac{{p_k{{\left| {\mathbf{h}_k^H\mathbf{F}_g^{}\mathbf{w}_k^{}} \right|}^2}}}{{\sum\nolimits_{\forall i \notin {\cal S}_g^{}}^{} {p_k{\lvert \mathbf{h}_i^H\mathbf{F}_g^{}{\mathbf{w}_k}\rvert}^2 + 1} }}.\begin{array}{*{20}{c}}
\end{array}
\vspace{-1mm}
\end{equation}
\indent
Since the RF precoder is optimized based on spatial channel correlation matrices, it is natural to consider the average SLNR $\mathrm{E}[\mathrm{SLNR}_k]$ as the performance metric. However, it is difficult to obtain the closed-form expression for $\mathrm{E}[\mathrm{SLNR}_k]$. Therefore, we derive a closed-form lower bound approximation of $\mathrm{E}[\mathrm{SLNR}_k]$ as follows.\\
\indent
Note that $\forall k\in \mathcal{S}_g$, $\mathbf{w}_k$ is independent of $\mathbf{h}_k$\footnote{This is because when the number of users $S_g$ is equal to the dimension of the effective channel $\overline{\mathbf{H}}_{g}$ in group g, the ZF precoding vector $\mathbf{w}_k$ for user $k$ is given by the unit vector along the direction orthogonal to the other users' channel vector $\mathbf{h}_j, j\neq k, j\in \mathcal{S}_g$. 
} and $\mathbf{h}_i, i\notin \mathcal{S}_g$, and $\mathrm{Tr}(\mathbf{F}_g\mathbf{F}^H_g)= S_g$\footnote{Without loss of generality, the magnitude of each element of $\mathbf{F}$ is assumed to be $\frac{1}{\sqrt{M}}$. Therefore, $\mathrm{Tr}(\mathbf{F}_g\mathbf{F}^H_g)=\frac{MS_g}{M}=S_g$.}. Assume that the spatial correlation matrices of the users in the same group are identical, i.e., $\mathbf{R}_k=\mathbf{R}_g,\forall k \in \mathcal{S}_g, \forall g\in \mathcal{G}$ \footnote{Here, $\mathbf{R}_k=\mathbf{R}_g$ is only used to provide a theoretical justification for the performance metric (SSLNR) of RF precoding design and similar assumptions can also be found in \cite{Liu2014Phase} and \cite{Adhikary2013Joint}.}, and $\mathrm{E}[\mathbf{w}_k\mathbf{w}^H_k]=\frac{\mathbf{I}_{S_g}}{S_g}$, which is true when $\mathbf{F}_g^H\mathbf{R}_g\mathbf{F}_g=\mathbf{I}_{S_g}$\footnote{In practice, $\mathrm{E}[\mathbf{w}_k\mathbf{w}^H_k]={\mathbf{I}_{S_g}}/{S_g}$ can be approximately satisfied when $\mathbf{F}_g^H\mathbf{R}_g\mathbf{F}_g$ has a good condition number (i.e., the ratio between the maximum and the minimum eigenvalues of $\mathbf{F}_g^H\mathbf{R}_g\mathbf{F}_g$ is not large compared to $1$).}. Then, the lower bound of the average SLNR of user $k$ in group $g$ can be approximated as
\vspace{-2mm}
\begin{small}
\begin{align}
&\!\!\!\!{{\mathop{\rm E}\nolimits} [{\rm{SLNR}}_k^{}{\rm{]}}\ge}\notag\\\label{Eq:90}
&\!\!\!\!\small{\frac{{{\mathop{\rm E}\nolimits} [{\bf{h}}_k^H{{\bf{F}}_g}{\mathop{\rm E}\nolimits} [{\bf{w}}_k^{}{\bf{w}}_k^H\left| {{\bf{h}}_k^{}} \right.]{\bf{F}}_g^H{\bf{h}}_k^{}]}}{{\sum\limits_{\forall i \in {{\cal S}_{g'}},g' \ne g}^{} {{\mathop{\rm E}\nolimits} [{\mathop{\rm Tr}\nolimits} ({\bf{h}}_i^H{{\bf{F}}_g}{\bf{w}}_k^{}{\bf{w}}_k^H{\bf{F}}_g^H{\bf{h}}_i^{})} ] + \frac{{K{\mathop{\rm Tr}\nolimits} ({{\bf{F}}_g}{\mathop{\rm E}\nolimits} [{\bf{w}}_k^{}{\bf{w}}_k^H]{\bf{F}}_g^H)}}{P}}}}\\ \label{Eq:91}
\vspace{-11mm}
&\small{\ge \frac{{\frac{1}{{{S_g}}}{\mathop{\rm Tr}\nolimits} \left( {{\bf{F}}_g^H{{\bf{R}}_g}{{\bf{F}}_g}} \right)}}{{\sum\nolimits_{g' \ne g}^{} {S_{g'}^{}{\mathop{\rm E}\nolimits} \left[ {{\mathop{\rm Tr}\nolimits} \left( {{\bf{w}}_k^{}{\bf{w}}_k^H} \right)} \right]{\mathop{\rm Tr}\nolimits} \left( {{\bf{F}}_g^H{{\bf{R}}_{g'}}{{\bf{F}}_g}} \right) + \frac{K}{P}} }}}\\
 \label{Eq:92}
&\small{ = \frac{{{\mathop{\rm Tr}\nolimits} \left( {{\bf{F}}_g^H{{\bf{R}}_g}{{\bf{F}}_g}} \right)}}{{\sum\limits_{g' \ne g}^{} {{S_g}{S_{g'}}{\mathop{\rm Tr}\nolimits} \left( {{\bf{F}}_g^H{{\bf{R}}_{g'}}{{\bf{F}}_g}} \right) + \frac{{K{S_g}}}{P}} }}\buildrel \Delta \over =\mathrm{SSLNR}_g,\forall k\in\mathcal{S}_g,}
\end{align}
\end{small}
\noindent where (\ref{Eq:90}) follows the Mullen's inequality \cite{Wang2012Statistical} and the inequality (\ref{Eq:91}) is due to $\mathrm{Tr}(\mathbf{A}\mathbf{B})\leq\mathrm{Tr}(\mathbf{A})\mathrm{Tr}(\mathbf{B})$\cite{Y2000A}. Note that all users in group $g$ share the same lower bound of the average SLNR and we call $\mathrm{SSLNR}_{g}$ in (\ref{Eq:92}) the statistical SLNR (SSLNR) of group $g$. It is expected that a good RF precoder can be found by maximizing the minimum SSLNR of all groups as follows:
\vspace{-1mm}
\begin{align}
&\mathcal{P}: \max\limits_{\{\mathbf{F}_g\}} \min\limits_{g\in \mathcal{G}} 
\mathrm{SSLNR}_g \notag\\
\vspace{-4mm}
\label{Eq:18}
s.t.&\quad f_{m,l} \in \{0,\frac{1}{\sqrt{M}},\frac{1}{\sqrt{M}}\emph{e}^{\mathrm{j}\frac{2\pi}{2^B}},\dots,\frac{1}{\sqrt{M}}\emph{e}^{\mathrm{j}\frac{2\pi (2^B-1)}{2^B}}\};\\  \label{Eq:19}
&\quad \sum\nolimits^{L}_{l=1}{\lvert f_{m,l}\rvert}^2={1}/{M}, \forall m\in \mathcal{M};\\
\label{Eq:88}
&\quad {1}/{\sqrt{M}}\leq\Vert \mathbf{f}_l\Vert\leq 1 ,l=1,\dots,L.
\end{align}
\indent
Assuming that $B$ bits quantized APSs are applied to the ACN, (\ref{Eq:18}) presents the feasible phases and $f_{m,l}=0$ means that no APS is selected between the $l$-th RF chain and the $m$-th antenna. Meanwhile, as restricted by the architecture in \textmd{Fig.}~\ref{fig:firure_1}, constraint (\ref{Eq:19}) restricts that each antenna is only connected to one RF chain through one APS, while constraint (\ref{Eq:88}) ensures that at least one APS is connected to the $l$-th RF chain.
\vspace{-2mm}
\subsection{Joint-Design for ACN and APSs (RF Precoder Optimization)}
\vspace{-1mm}
\indent 
Problem $\mathcal{P}$ is very challenging since it involves both the non-convex objective and constraints as well as discrete optimization variables. To tackle these challenges, we resort to a heuristic algorithm called the GRFP algorithm, which first finds the optimal RF precoder with relaxed constraints, then approximately ``projects'' the relaxed solution to the constraint set of $\mathcal{P}$ by properly choosing the ACN connection and analog phases. Consider the following relaxed problem of $\mathcal{P}:$ 
\vspace{-1mm}
\begin{equation*}
{\widetilde{\mathcal{P}}}:\max\nolimits_{\{\mathbf{F}_g\}} \min\nolimits_{g\in \mathcal{G}} 
\mathrm{SSLNR}_g \quad s.t.\quad (\ref{Eq:88}) \notag.
\vspace{-2mm}
\end{equation*}
${\widetilde{\mathcal{P}}}$ can be decomposed into $G$ independent subproblems as
\vspace{-1mm}
\begin{equation}
{\widetilde{\mathcal{P}_g}}: \max \nolimits_{\mathbf{F}_g} \mathrm{SSLNR}_g 
\vspace{-1mm}
\label{67}
\quad s.t.\quad{1}/\sqrt{M}\leq\Vert \mathbf{f}_l\Vert\leq 1 ,l\in \mathcal{L}_g.
\vspace{-1mm}
\end{equation}
For $g=1,\dots,G$, ${\widetilde{\mathcal{P}_g}}$ is equivalent to the following problem:
\vspace{-1mm}
\begin{equation}
\widehat{\mathcal{P}_g}: \quad \max \limits_{\alpha_g,\mathbf{F}_g} \alpha_g 
\quad s.t.\quad \mathrm{SSLNR}_g\geq \alpha_g, (\ref{67}). 
\vspace{-2mm}
\end{equation}
Then, the constraint $\mathrm{SSLNR}_g\geq \alpha_g$ can be rewritten as
\vspace{-2mm}
\begin{equation}
\mathrm{Tr}(\mathbf{R}_{g}\mathbf{F}_{g} \mathbf{F}_{g}^H)-\alpha_g\mathrm{Tr}(\overline{\mathbf{R}}_{g} \mathbf{F}_{g} \mathbf{F}_{g}^H)\geq \frac{KS_g }{P}\alpha_g ,
\vspace{-2mm}
\end{equation}
where $\overline{\mathbf{R}}_{g}=\sum_{g'\neq g} S_gS_{g'} \mathbf{R}_{g'}$. Since $\widehat{\mathcal{P}_g}$ is still non-convex, $\alpha_g$ and $\mathbf{F}_g, \forall g\in \mathcal{G}$ are iteratively solved using the following process. Given $\alpha_g$, the optimal solution to $\widehat{\mathcal{P}_g}$ can be obtained by solving 
\vspace{-2mm}
\begin{equation}\label{Eq:777}
f(\alpha_g)\buildrel \Delta \over = \max\nolimits_{\mathbf{F}_g}\{\mathrm{Tr}(\mathbf{R}_g \mathbf{F}_g \mathbf{F}^H_g)-\alpha_g\mathrm{Tr}(\overline{\mathbf{R}}_g\mathbf{F}_g \mathbf{F}^H_g)\}.
\end{equation}
Let $\mathbf{R}_g-\alpha_g\overline{\mathbf{R}}_g=\overline{\mathbf{U}}_g(\alpha_g) \mathbf{D}_g(\alpha_g) \overline{\mathbf{U}}_g^H(\alpha_g)$ denote the eigenvalue decomposition of $\mathbf{R}_g-\alpha_g\overline{\mathbf{R}}_g$, where the diagonal elements of $\mathbf{D}_g(\alpha_g)$ are in descending order. Then, the optimal solution to (\ref{Eq:777}) is
\vspace{-1mm}
\begin{equation}\label{68}
\mathbf{F}^*_g(\alpha_g)=[\widetilde{\mathbf{U}}_g(\alpha_g)\boldsymbol{\Lambda}_{g}\left(\alpha_{g}\right)],
\vspace{-1mm}
\end{equation}
where $\widetilde{\mathbf{U}}_g(\alpha_g)\in \mathbb{C}^{M\times {S}_g}$ consists of ${S}_g$ dominant eigenvectors of $\overline{\mathbf{U}}_g(\alpha_g)$ corresponding to ${S}_g$ largest eigenvalues in $\mathbf{D}_g(\alpha_g)$, $\boldsymbol{\Lambda}_{g}\left(\alpha_{g}\right)$ is a diagonal matrix with the $i$-th diagonal element given by $\Lambda_{g}^{i}\left(\alpha_{g}\right)=\begin{cases}
1 & if\:d_{g}^{i}\left(\alpha_{g}\right)\geq0\\
\frac{1}{\sqrt{M}} & otherwise
\end{cases}
$, and $d_{g}^{i}\left(\alpha\right)$ is the $i$-th largest eigenvalue in $\mathbf{D}_{g}\left(\alpha_{g}\right)$. Then, the corresponding optimal value of problem $\widehat{\mathcal{P}_g}$ is given by
\vspace{-1mm}
\begin{equation}\label{69}
f(\alpha_g)=\sum\nolimits_{i=1}^{{S}_g}d_g^i (\alpha_g)\Lambda_{g}^{i}\left(\alpha_{g}\right).
\end{equation}
Finally, the maximum feasible $\alpha_g$, denoted by $\alpha_g^*$, can be found by solving the following equation using the bisection method:
\vspace{-2mm}
\begin{equation}\label{81}
f(\alpha_g)=\frac{KS_g }{P}\alpha_g.
\vspace{-3mm}
\end{equation} 
The corresponding optimal RF precoder for ${\widetilde{\mathcal{P}}}$ is $\mathbf{F}_g^*=\mathbf{F}^*_g(\alpha_g^*)$.\\
\indent
However, the obtained $\mathbf{F}^*_g$ does not satisfy the discretization of the analog phases in (\ref{Eq:18}) and the antenna structure limitations in (\ref{Eq:19}), (\ref{Eq:88}). Based on the relaxed solution $\mathbf{F}^*_g$, we use a greedy method to assign the $M$ available APSs to form the $M$ non-zero elements of $\mathbf{F}$. Details are found in the following \textbf{Algorithm 1}. 
\vspace{-3mm}
\begin{algorithm}\scriptsize\label{algorithm2}
\caption{Group RF Precoding (GRFP)}	
\textbf{Input:} APS quantized bits $B$, spatial correlation matrices $\{\mathbf{R}_g\},\forall g\in \mathcal{G}$, user groups set $\mathcal{S}_g$, group quantity $G$ and antenna quantity $M$.\\	
\textbf{Output:} RF precoding matrix for the $g$-th group $\mathbf{F}_g$.\\
$1:$ $\forall g \in \mathcal{G}, \mathbf{F}_g=\mathbf{0}$.\\
$2:$ Calculate $\alpha_g^*,\mathbf{F}_g^*\in \mathbb{C}^{M\times {S}_g}$ for group $g,\forall g\in \mathcal{G}$ according to (\ref{68})-(\ref{81}).\\
$3:$ Sort $\alpha_g^*,\forall g\in \mathcal{G}$ in ascending order and denote the index of the sorted group as $t(j), j=1,\dots, G$.\\
$4:$ $\mathcal{X}=\varnothing$;\quad count=0;\quad flag=false;\\
\For{$j=1$ to $G$} 
{
$g=t(j);$\\
			\For{$i=1$ to $S_g$} 
			{
			$m^*=\mathop{\argmax}_{m\in \mathcal{M}/\mathcal{X}}\{\lvert f^*_{m,i}\rvert \},
				s.t. \sum^{L}_{l=1}\lvert f_{m,l}\rvert=0$,where $f_{m,i}^{*}$ denotes the element in the $m$-th row and the $i$-th column of $\mathbf{F}^*_g$; \\
			$\mathcal{X}= \mathcal{X}\cup \{m^*\}$;\\
			$n^*=\mathop{\argmin}_{n\in \{0,1,\dots,{2^B-1}\}}\left\vert \dfrac{f_{m^*,i}^{*}}{\lvert f_{m^*,i}^{*} 				\rvert}-\emph{e}^{\mathrm{j}\frac{2\pi n}{2^{B}}}\right\vert$;\\
			$f_{m^*,i}=\emph{e}^{\mathrm{j}\frac{2\pi n^*}{2^{B}}}$,
			count++;\\
			\textbf{if} count == M \, flag=true;\,\, \textbf{break};\\
			
			}

	\textbf{if} flag == true \textbf{break};\\
	\textbf{end}\\
}
\quad Obtain the RF precoding matrices $\{\mathbf{F}_g\},\forall g\in \mathcal{G}$.
\end{algorithm}
\vspace{-2mm}
\begin{figure*}
\centering  
\makeatletter\def\@captype{figure}\makeatother 
\subfigure[]{    \label{fig_2}
\hspace{-4mm}      	
	\begin{minipage}{7cm}
		\centering  
		\includegraphics[width=6.4cm, height=4.2cm]{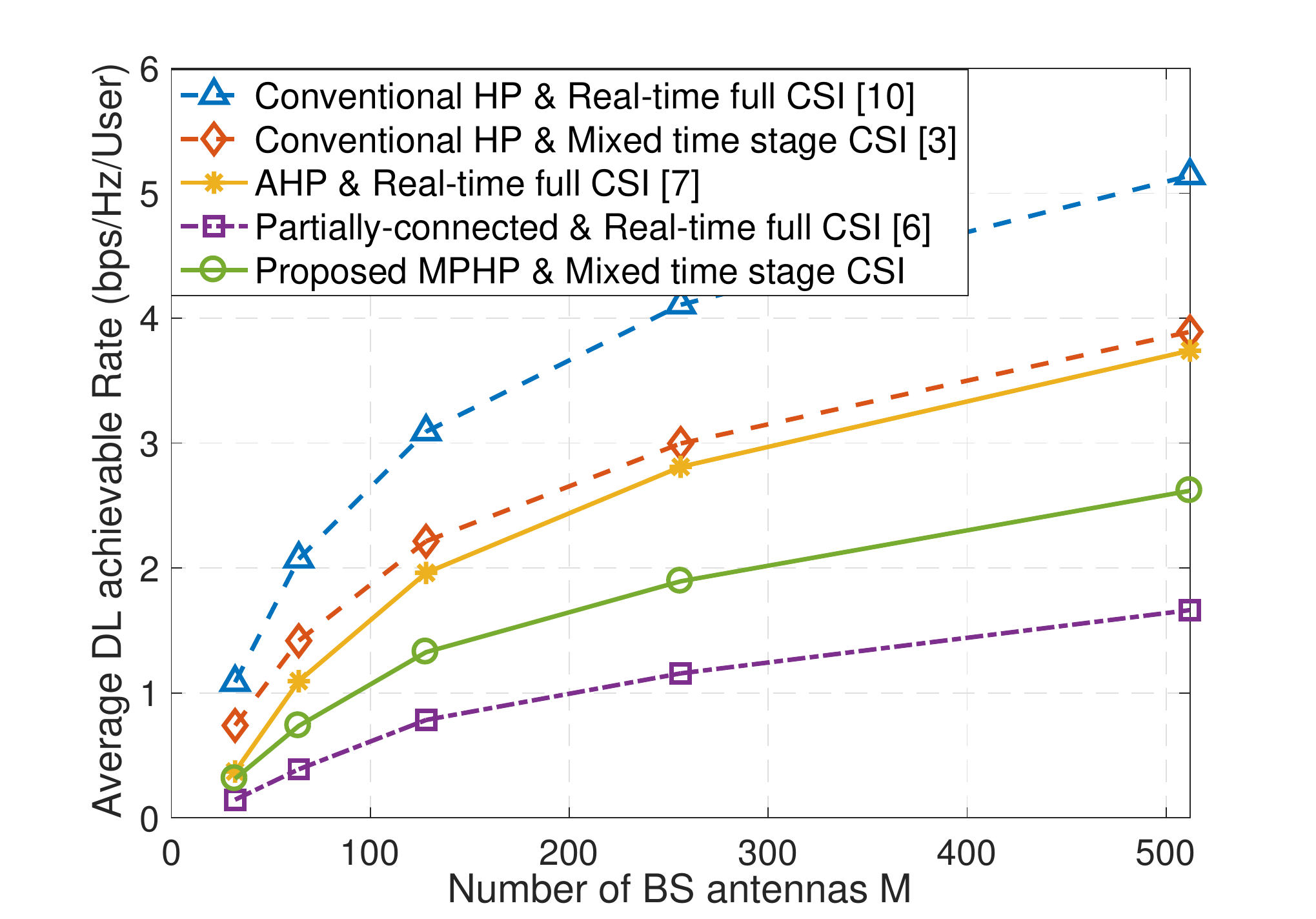}\label{fig:firure_2}
	\end{minipage} }
	\hspace{-14mm}  
\subfigure[]{    \label{fig_3}    
    \begin{minipage}{7cm}
    	\centering  
		\includegraphics[width=6.4cm, height=4.2cm]{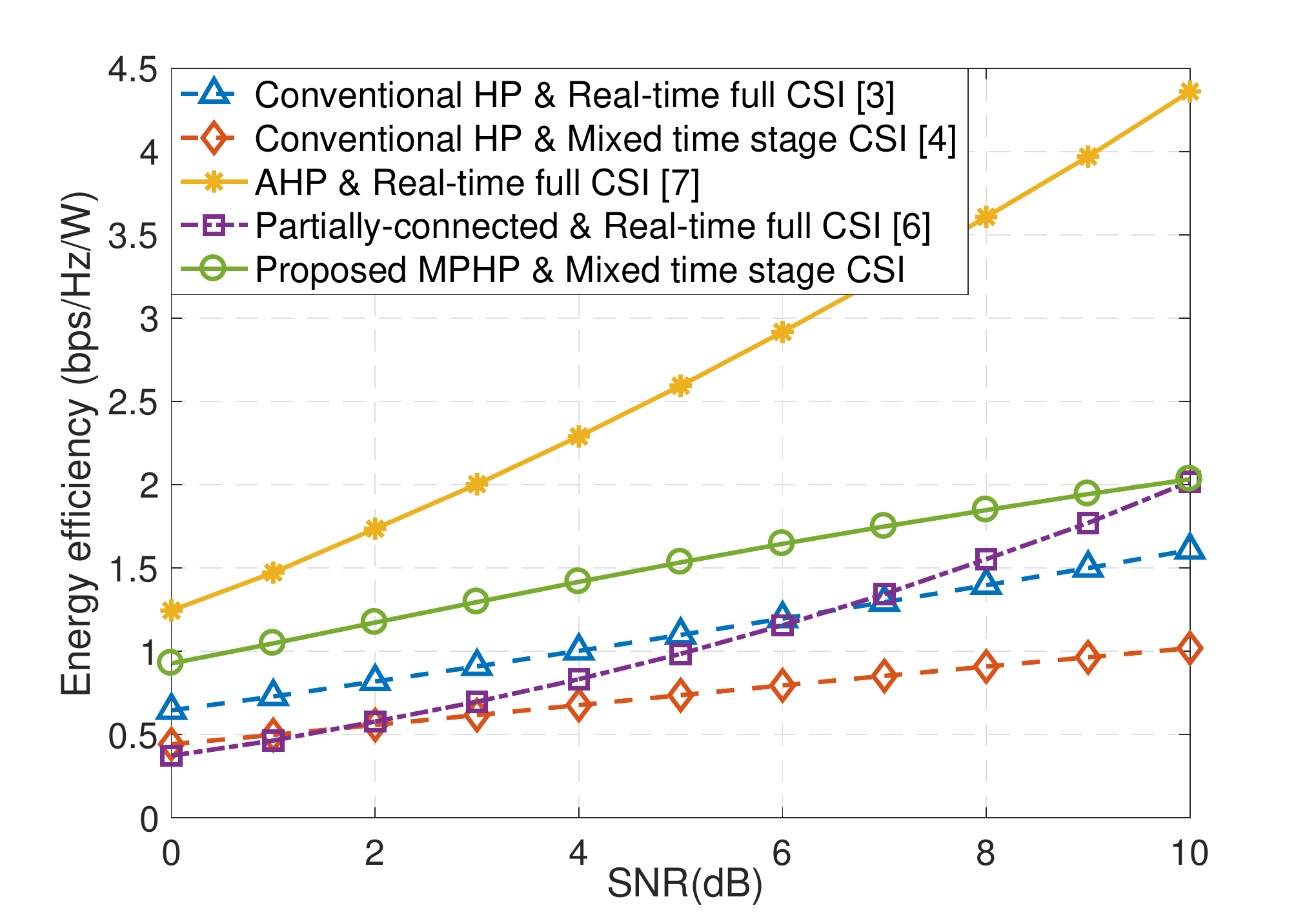}\label{fig:firure_3}
	\end{minipage}} 
\hspace{-13mm}  
\subfigure[]{    \label{fig_4}     	
	\begin{minipage}{7cm}
		\centering  
		\includegraphics[width=6.4cm, height=4.2cm]{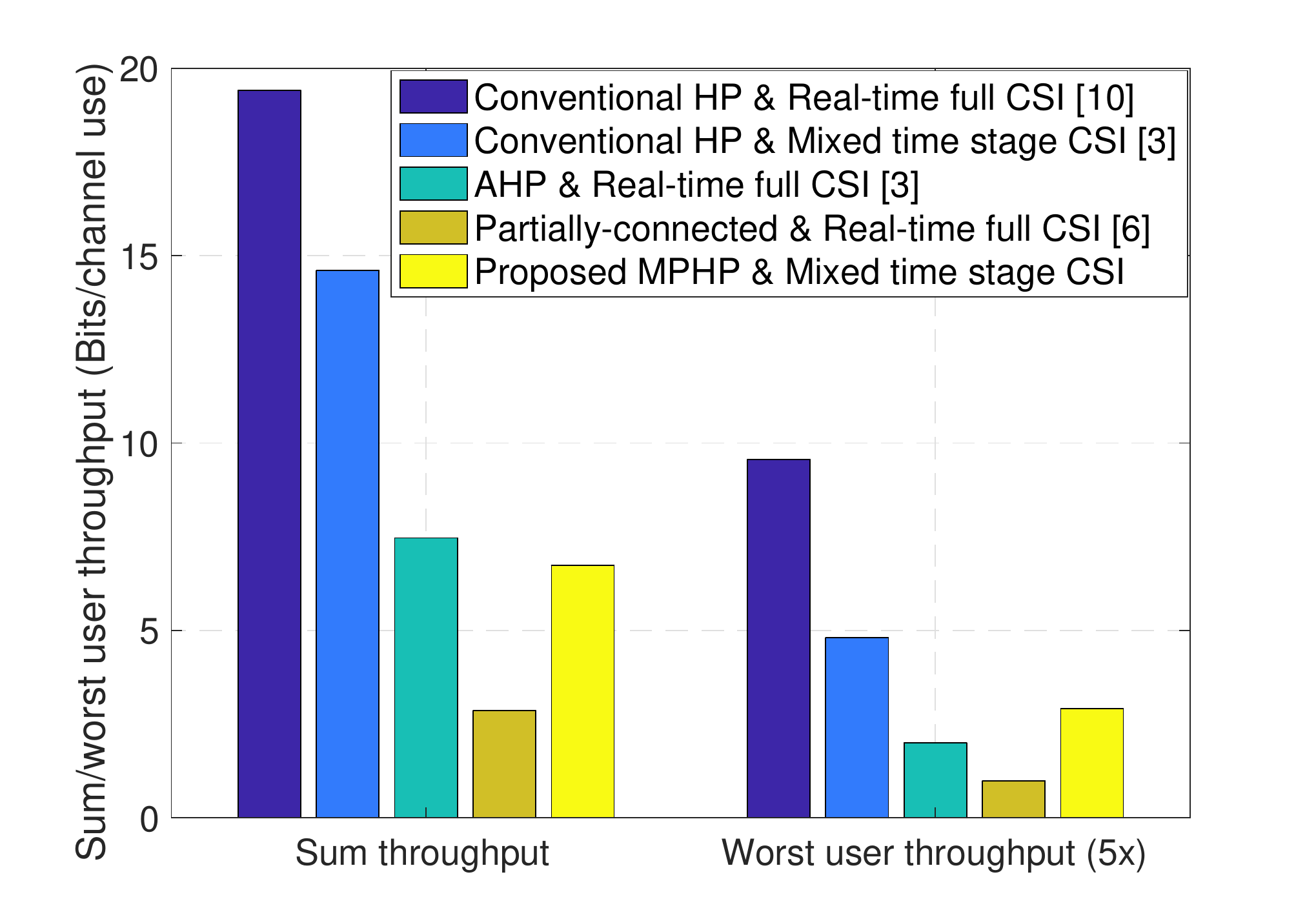}\label{fig:firure_4}
	\end{minipage} }
\caption{Simulation results: (a) The average achievable rate per user against the number of BS antennas $M$ with $B=4$; (b) The comparison of energy efficiency with $M=64$ and $B=4$; (c) Throughput comparisons for different schemes with  $M=64$ and $B=4$.} \label{fig_two feasible timetables} 
\end{figure*} 
\vspace{-6mm}
\section{Simulation Results}\label{SimuRe}
\vspace{-1mm}
In this section, numerical simulations are presented and compared with the conventional HP using real-time full CSI\cite{Liang2014Low}, the conventional HP with mixed time stage CSI\cite{Liu2014Phase}\footnote{Here, mixed time stage CSI refers to the knowledge of spatial correlation matrices plus the effective CSI of all users. 
}, the AHP utilizing the real-time full CSI\cite{Zhu2016Adaptive}, and the HP scheme under the PRPS utilizing the real-time full CSI \cite{6824962}. A typical downlink of an FDD multi-user massive MIMO system is considered. There are $8$ users distributed in a circular cell over a distance of $[35,500]$ meters. The downlink channel is generated using the spatial channel model (SCM) in 3GPP\cite{WinnerScmImplementationIEEETranbst}. Users are partitioned into $3$ groups using the grouping algorithm in \cite{Xu2014User} according to the similarity of the spatial correlation matrices. The total transmit power at the BS is defaulted as $P=1\,\mathrm{W}$ and $L=K$.\\
\indent
\textmd{Fig.}~\ref{fig:firure_2} plots the average achievable rate per user against $M$. The achievable rate increases with the number of antennas at a similar rate. The conventional HP with real-time full CSI in\cite{Liang2014Low} exhibits the highest average achievable rate, while with statistical CSI, the rate is slightly lower for both FRPS and adaptive PRPS due to lack of instantaneous CSI. In contrast to the baseline partially-connected \& real-time full CSI, the proposed scheme achieves higher per-user achievable rate due to the selection gain by the ACN.
\textmd{Fig.}~\ref{fig:firure_3} presents EE comparison, where $EE=SR/(P+P_{\mathrm{BB}}+LP_{\mathrm{RF}}+N_{\mathrm{APS}}P_{\mathrm{APS}})$ and the value of each parameter defined accords to that in \cite{M2015Hybrid}. Note that the total power consumption depends on the number of required APSs $N_{\mathrm{APS}}$, which is $ML$ for the FRPS, and $M$ otherwise. It can be seen that the EE of the proposed HP scheme is much higher than that of the two baseline schemes under the FRPS, which reveals that the adaptive PRPS is more efficient in energy cost. As for the partially-connected \& real-time full CSI method, although it exhibits the similar hardware cost, its EE is much lower than the proposed scheme when the signal-to-noise ratio (SNR) is less than 10 dB, which is the typical SNR region in practical massive MIMO systems. Note that this baseline achieves a higher EE when SNR is larger than 10 dB because it benefits more from the real-time full CSI for higher SNR.
In \textmd{Fig.}~\ref{fig:firure_4}, we show the sum and worst user throughput comparisons of different schemes. The results show that the conventional HP under the FRPS can achieve a better throughput performance. The worst user throughput of the proposed MPHP is higher than that of the AHP\cite{Zhu2016Adaptive} and the partially-connected scheme\cite{6824962}. Moreover, the Jain's fairness index of the proposed MPHP's is $(\sum_{k=1}^KR_k)^2/(K\sum_{k=1}^KR_k^2)=0.9518$, while that of AHP is $0.8202$. Furthermore, the comparisons of the computational complexity and CSI feedback overhead are presented in \textmd{TABLE}~\ref{tab2}.
\vspace{-2mm}
\begin{center}
	\begin{tablehere}
			\vspace{-1mm}
		\scriptsize
		\caption{The Comparison of Algorithms}\label{tab2}
			\vspace{-3mm}
		\begin{tabular}{|c|c|c|}
			\hline
			 & Computational Complexity & Feedback\\
			\hline
			AHP \& real time CSI & higher & $ MKT $\\
			\hline
			MPHP \& mixed time stage CSI & lower & $T\sum_{g} S_g^2+Z $\\
			\hline
		\end{tabular}
	\end{tablehere}	
\end{center}
	\vspace{-1mm}
Let $T$ denote the coherence time of the channel statistics. The proposed MPHP only updates the RF precoder once for $T$ time slots with the knowledge of $\mathbf{R}_g, \forall g\in\mathcal{G}$, while the short-term precoder is calculated for each time slot with the reduced-dimension CSI $\overline{\mathbf{H}}_{g}, \forall g\in\mathcal{G}$. The feedback overhead of $\overline{\mathbf{H}}_{g}$ and the spatial correlation matrices for $T$ time slots are $TS_g^2$ and $Z$, respectively, where $Z\leq KM^2$ (this is because we only need to feedback the dominant eigenvalues and eigenvectors of each covariance matrix). However, the AHP \& real-time stage CSI requires the instantaneous CSI $\mathbf{H}$ with feedback $MKT$. Therefore, the longer $T$ is, the lower the computational complexity and feedback that can be achieved by the proposed MPHP.\\
\indent
In summary, the proposed MPHP has a higher EE than the conventional HP in\cite{Liu2014Phase},\cite{Liang2014Low} and the HP scheme in\cite{6824962}, and a lower complexity and feedback overhead than the conventional HP, AHP, and the HP scheme in\cite{6824962} utilizing the real-time full CSI. Therefore, the proposed MPHP can achieve a good tradeoff between the performance, complexity and CSI signaling overhead, which makes it an attractive solution in practice.
\vspace{-2mm}
\section{Conclusions}\label{Conclu}
\vspace{-1mm}
In this letter, a mixed-timescale per group HP structure is proposed to reduce the CSI signaling overhead under the adaptive PRPS in multi-user massive MIMO systems. A GRFP algorithm is also proposed to solve the RF precoder optimization problem under the MPHP structure. 
Future work can be done on designing a more efficient grouping algorithm for the MPHP to further improve the performance.
\bibliographystyle{IEEEtran}

\end{document}